\documentclass[12pt]{article}
\usepackage{epsf}
\usepackage{latexsym}

\begin{document}

 \title{ RANDOM ROUGHNESS OF
BOUNDARY INCREASES TURBULENT CONVECTION SCALING EXPONENT}
 \author{ S. Ciliberto, C. Laroche \\
         Ecole Normale
Sup\'erieure de Lyon, Laboratoire de
Physique ,\\
 C.N.R.S. URA1325,  \\ 46, All\'ee d'Italie 69364 Lyon Cedex
07 France\\
        }
\maketitle

%
%
%
%

\begin{abstract}

The influence of the boundary layer properties on the heat transport in
 turbulent thermal  convection is
 experimentally investigated in a cell with a rough bottom plate.
It is shown that the standard 2/7 exponent of the
 convective heat flow  dependence on the  Rayleigh number, usually observed
 in cell with smooth boundaries,
increases if the roughness
has power law distributed asperity heights and the thermal
boundary layer thickness is smaller than the maximum asperity size.
 In contrast  a periodic roughness does not influence  the  heat
transport law exponent.

\end{abstract}

\medskip

\newpage

In the last years many models and experiments
have  been done in order to understand
the heat transport properties of
turbulent thermal convection in a fluid layer heated from below,
 that is
Rayleigh Benard convection
\cite{SigII,Cast,Chilla,Chav,Siggia,Lohse,Toschi}.
These properties are characterized by the dependence on
the Raleigh number $Ra$
of the non-dimensional heat flow, that is the Nusselt number
 $Nu$. In many experiments it has been observed that, for $Ra>10^6$,
 $Nu$ has a power law dependence on $Ra$, that is:
$Nu = \alpha Ra ^\gamma $. Specifically,
in fluids with Prandtl number of
about 1, $\alpha$ and $\gamma $ take the following values for
$10^6<Ra<10^{11}$:   $\gamma=2/7$ and $\alpha\simeq 0.2$
\cite{Cast,Chilla,Chav}.

In order to construct a reliable model of the  heat transport law
several aspects of the turbulent flow  have to be solved. For
example it is still unclear how  the heat transport is influenced
by  the size distribution of thermal fluctuations   in the
boundary layer and by the
 coupling of these fluctuations with
the mean circulation flow (MCF). Thermal fluctuations of the
boundary layer are associated with   thermal plumes
\cite{HWK,Cast} and the MCF  is
 a large scale convective roll involving  all the cell
containing the convective fluid
 \cite{HWK,Sano}.

The role of MCF on the heat flux  has been studied in several experiments,  but
 the values of  $\gamma$ and $\alpha$ are not modified by
either  the perturbation
\cite{Xia,Gollub} or the  suppression \cite{Cili} of the MCF.  No influence on
the value  of $\gamma$
 has been observed in numerical
simulation
where no-slip boundaries were used \cite{Benzi}.
In two recent experiments \cite{Shen, Du} boundaries with periodic roughness
have
been used. In this case
$\alpha$  is much  larger than the value measured in cell with smooth
boundaries, that is the heat flux is enhanced. However it is important to stress
that
a periodic roughness  does not modify the value of  $\gamma$ which is still
  $2/7$ as in  the case of smooth  boundaries.
Finally it is worth to mention that
in a recent experiment the
 transition toward the ultimate regime  has been observed \cite{Chav}.
 This transition, which manifests itself with an increasing of
$\gamma$ for $Ra>10^{11}$,  has been explained by the change of
the dissipation properties in the thermal and viscous boundary
layers (see also ref.\cite{Toschi}). One of the consequences that
one may extract from all these  experiments and simulations is
that the value of $\gamma$ could be mainly controlled by thermal
fluctuations (the thermal plumes) and by their size distribution
in the boundary layer.  Therefore a perturbation of this
distribution may change the value of $\gamma$.

 The   purpose of this letter is to show that using a very rough boundary,
 with  power law distributed asperity heights, it is possible to strongly
 modify the value of $\gamma=2/7$ observed in cell with smooth boundaries.
The influence of the boundary roughness  on transport properties
 is   a very important topic in turbulence \cite{Schl,Morn}. This
topic
 has not been widely studied in turbulent thermal convection. In
refs.\cite{Shen,Du} only a periodic roughness  has been
considered. As already mentioned, this kind of roughness does not
modify the value of $\gamma$ but only that of $\alpha$. Therefore,
in this letter we want to  stress  the difference between a
periodic and a random roughness.

The experimental  apparatus has been already described in ref.\cite{Chilla}
and  we recall here only the main features.
  The cell has horizontal sizes $L_x=40 cm$ and $L_y=10 cm$
and
      two different  heights $d$ equal to $20 cm$ and $10 cm$.
        With these two
cells filled with water (at an average temperature  of
         $45^o  C$ corresponding to a Prandtl number,   of
         about  3) we are able to cover the interval
         $10^7 < Ra < 10^{10}$.
         The  bottom copper  plate
         plate is  heated with an electrical resistor. The top copper  plate
     is     cooled by a water circulation and  its  temperature is
         stabilized by an electronic controller. All the apparatus
        is  inside a temperature stabilized box.
 The temperature of the  plates is measured
in several locations.
Local temperature
measurements, of the turbulent flow,
are done with two small
thermocouples (P1, P2)
of $0.04 cm$ in diameter with a response time of $5ms$.
The probes P1 and P2 are   located at $(L_x/4,L_y/2)$ and
   at ($L_x/2,L_y/2$) respectively.
Both probes can be moved  along z with micrometric devices in
order to measure the mean temperature profile and that of the
temperature fluctuations as a function of $Ra$. To measure the
heat flow we first estimate the heat losses of the cell
\cite{Chilla}. These heat losses are then subtracted from the
heating power to evaluate the fraction of heat effectively
transported by the convective water.

 We  perturb the bottom boundary layer by changing  the
roughness of the bottom plate. This  roughness is  made by  small
glass spheres of controlled diameter glued on the bottom copper plate,
 with a very
thin layer of thermal  conductive  paint.
 We use
 $N$ sets of spheres such that each set $j$ is  composed by spheres
  having  the same diameter $h_j$, with $1  \le j  \le N$
  and $h_1 < h_2 ....... < h_N$.
The  sphere number  $P(h_j)$ in each set
is selected in order to
produce a well defined  power law distribution,
that is $P(h_j) = A  \ h_j^{-\xi}$.
Here $A$ is a normalization factor such that the ensemble of  spheres
covers uniformly the copper plate surface, that is
 $ L_x \times  L_y = {\pi/4}  \sum_{j=1,N} h_j^2 P(h_j) $.
The spheres are mixed and randomly glued on  the bottom plate.
 The roughness properties
  can  be  changed
 by modifying $N$, $\xi$, the minimum sphere diameter $h_1$  and
 the maximum sphere diameter $h_N$.

 The roughness has two important characteristics lengths
 $h_1$ and $h_N$.
 These  lengths have  to be compared with one of the main characteristic
 length of turbulent thermal convection that is the thermal boundary
 layer thickness $\lambda=d/2/Nu$. Indeed if $\lambda >> h_N$
 or $\lambda < h_1$ it is reasonable to think that no effect on the
 convection thermal properies will  be observed.
 In contrast if $h_1<\lambda<h_N$  several important
 changes could be produced. The mean roughness height $\tilde h$
does not seem to  play any important role.

In order to understand the role played by the roughness
on the boundary layer,
 we performed several experiments.  In four experiments,
  labeled I,II, III,IV respectively, the surface roughness had
  power law distributed asperities.
 A periodic roughness,  with only one characteristic length, was used
  in another  experiment labeled V.
 Specifically
the roughness parameters, in the different experiments,
 took the values indicated in Table I.
The results of these five experiments have been compared to those
obtained in the same cell with smooth boundaries \cite{Chilla,Cili}.

The non-dimensional convective  heat flow $Nu$
versus $Ra$ measured in a cell
with smooth boundaries is compared in fig.1
with that measured in the experiments (III and IV).
We clearly see that in the three  cases
 $Nu$ is a power law function of $Ra$,
that is $Nu\propto Ra^\gamma$ with $\gamma\simeq2/7$
in the smooth plate case, $\gamma \simeq 0.45$ in
experiment IV ($\xi=1$)
and $\gamma=0.35$ in experiment III($\xi=2$).
It is important to notice that in the experiments III
 and IV the thermal boundary
layer thickness $\lambda$ is always smaller than
$ h_N$ in all the interval $10^8< Ra<10^{9.5}$
where these two experiments have been performed.
Indeed from fig.1 one sees that
at $Ra=10^8$,  $Nu \simeq 16$, therefore $\lambda =0.63 cm< h_N$
in experiments III and IV.
These measurements clearly show that,
when $\lambda<h_N$ has a value comparable to that of the roughness,
the dependence of $Nu$ as a function of $Ra$ is strongly
modified and $\gamma$
is a function of $\xi$,  that is the exponent of the roughness height
distribution $P(h)$. Specifically $\gamma$ increases
when the roughness height distribution becomes more flat.

In order to show that
for $\lambda> h_N$ no effect on
Nu is observed we describe  the results
of the experiments (I, II).
These two experiments have the same $\xi$ of experiment III
but the important difference is that
$\lambda$ becomes  comparable to   $h_N$
in the middle of the  $Ra$ spanning range.
Specifically from the $Nu$ measured in these experiments,
 plotted as  a function of $Ra$ in figs. 2a) and 2b),we find
 $\lambda \simeq 0.2 cm$ at $Ra \simeq 1.5 \ 10^{9}$ for experiment I
and $\lambda=0.2 cm$ at $Ra\simeq 3 \ 10^{8}$ for experiment II.
In fig.2a) we clearly see that when $\lambda< h_N$, that is for
$Ra\ge  10^{9}$ the dependence of Nu changes and we find
$\gamma=0.35$ as in experiment  III. The value of $\gamma$ in this
figure and in the next is certainly  not very precise because of
the very limited scaling range. However the exact value of
$\gamma$ is not very important for the discussion. What we want to
show here is just the clear change of trend for  $Ra\ge  10^{9}$.
In contrast for $Ra <  10^{9}$, that is for $\lambda>h_N$, we see
that the experimental points are parallel to those corresponding
to the smooth plate. In fig.2b) the results of experiment II are
directly compared with those of experiment I. Experiment II has
the same roughness of experiment I and  a smaller d therefore
$\lambda< h_N$ at  $Ra\simeq 3 \ 10^{8}$. Indeed we see that $ Nu$
begins to increase faster for  $Ra>  3 \ 10^{8}$. At the same time
we notice that in both experiments I and II all the  points for
which $\lambda>h_N$ are  aligned on the same straight line
parallel to that of  the smooth case.
 Thus we see that in order to have an interaction
of the thermal boundary layer with  the roughness
$\lambda$ should be smaller than $h_N$.

To show that this is actually the case we have measured the profile of
the temperature and of the temperature fluctuations as  a function of z in
experiment I for two values of $Ra$. These profiles are plotted
in fig.3
 as a function of $0.5 z/\lambda= z Nu/d$. For $\lambda > h_N$
 the profile with roughness is  very close to that  with a smooth plate.
In contrast for  $\lambda < h_N$
 the dependence as a function of z of the temperature
 fluctuation rms and of temperature is
 strongly perturbed by the presence of the
 roughness, which induces  the appearance  of a second maximum in the
 rms profile. The position of this maximum is close to $h_N$.
 Thus the profile shape confirms that
 the dependence of $Nu$ as a function of $Ra$
 is modified by the roughness only when $ h_N > \lambda$.

 Finally we compare the results of the experiments I,..,IV
with those of experiment V which has a periodic roughness. The
curve $Nu$ versus $Ra$ measured in the experiment (V) is plotted
in fig.4. The presence of a jump in the curve is clearly observed.
The location of the jump corresponds to the  value of $Ra$ where
$\lambda \simeq h_N=h_1$. However above and below this jump the
slope of the curve is very close to that with smooth plates. These
results agree with those of ref\cite{Shen,Du} where a periodic
rough plate was used. They also tell us that when $\lambda<h_1$
the roughness does not influence  the value of $\gamma$. From the
comparison of the results of experiment V with those of experiment
III and IV one deduces that an important ingredient  for modifying
the exponent $\gamma$ is the presence of a roughness with a power
law distributed height and that a periodic roughness does not
influence the value of $\gamma$.

In spite of these important changes in the behaviour of $Nu$
versus $Ra$ in experiments I,II,III and IV  the bulk properties
are not modified by the presence of the roughness. The histograms
and spectra of the local temperature fluctuations measured in the
center of the cell by probe P2 and on the side by  probe P1 are
the same with and without  roughness. Furthermore the frequency
$f_c$ of the slow oscillation which is related to the MCF period,
has the same dependence on $Ra$ observed in  experiments with
smooth plates  and  $Pr \simeq O(1)$ \cite{Sano,Cili,Chav}. We
find that  our data are compatible with a law $f_c \simeq \chi
/d^2 \ A_f \ Ra^{0.49}$ with $A_f \simeq 0.06$ without roughness
and $A_f\simeq 0.05 $ with roughness. This result is in agreement
with those of ref.\cite{Chav}
 where it is shown
that $f_c$ is always proportional to $ Ra^{0.49}$ even for
$Ra>10^{11}$ where
  $\gamma>2/7$ as in our experiment  ref.\cite{Chav}. Therefore our experiment and
that of   ref.\cite{Chav} seems to indicate that there is a
negligible influence of the value of $\gamma$ on the dependence of
$f_c$ on $Ra$ and on the statistical properties.

At the moment we are unable to construct a model which explains
the dependence of $\gamma$ on $\xi$. The prediction of a  recently
proposed model does not agree with our observations
\cite{Vill}because this model predicts an increasing of $\gamma$
for increasing $\xi$ and we observe just the contrary.
Nevertheless one can try to understand why the presence of a
random roughness is so important. From refs.\cite{Xia,Cili}
 we know that
any perturbation of the MCF does not change the heat transport.
Furthermore several models do not need to rely upon the MCF in order
to explain the $Nu\propto Ra^{2/7}$ law \cite{Cast,Toschi}. Therefore
if one assumes that the heat
transport is mainly  due to plumes, whose characteristic size is close to
$\lambda$, then  if $\lambda>>h_N$
it is obvious that no perturbation of the heat transport is observed
and everything goes as in the smooth case.
In contrast when $h_1<\lambda<h_N$ and the roughness has a power law
distribution, the plumes and the boundary layer cannot construct their
characteristic length  because the roughness contains many different
lengths. As a consequence the modified
$\lambda$ changes the dependence of $Nu$ versus $Ra$.
Finally if  $\lambda<h_1$ no influence on $\gamma$ is observed
because the system is locally equivalent to a smooth  one.
It is also important to recall that
 in  ref.\cite{Chav}  the increase of
$\gamma$  for $Ra > 10^{11}$ has been justified by the change of
the dissipation properties in the boundary layer. From  all of
these observations
 one may extract the very important conclusion that
the turbulent heat transport is dominated by the thermal
fluctuations close to the boundary layer and by their size
distribution. If this distribution is perturbed by the presence of
a random roughness  the transport properties are modified too. Any
realistic model of thermal
 convection should take into
account these results, which  could be very useful
 in the study
of convection in  geophysical flows where the presence
of  smooth plates is certainly
a very idealized case.

We acknowledge very useful discussion with R. Benzi, D. Lohse,   J.F. Pinton and
E. Villermaux.

\begin{table}
\begin{center}
\begin{tabular}{|c|c|c|c|c|c|c|}
\hline
 Experiment & $P(h_j)$  & $N$
 & $h_1 [cm]$ & $h_N [cm]$ & $\tilde h [cm]$ & $d [cm]$ \\
\hline
I & $h^{-2}$ &3  & 0.06 &  0.4 & 0.08 & 20 \\
\hline
II &  $h^{-2}$ & 3 & 0.06 & 0.4 & 0.08 & 10 \\
\hline
III &  $h^{-2}$ & 5 & 0.06 & 1 & 0.08 & 20 \\
\hline
IV &  $h^{-1}$ & 5 & 0.06 & 1 & 0.08 & 20 \\
\hline
V & $\delta(h-h_N)$ & 1 & 0.2 & 0.2 & 0.2 & 20 \\
\hline
\end{tabular}
\end{center}
\caption{Roughness parameters: experiments I,II,III
 and IV have power law
distributed asperity heights whereas experiment
 V has a periodic  roughness}
\end{table}


\newpage


\begin{figure}
\centerline{\epsfysize=0.7\linewidth \epsffile{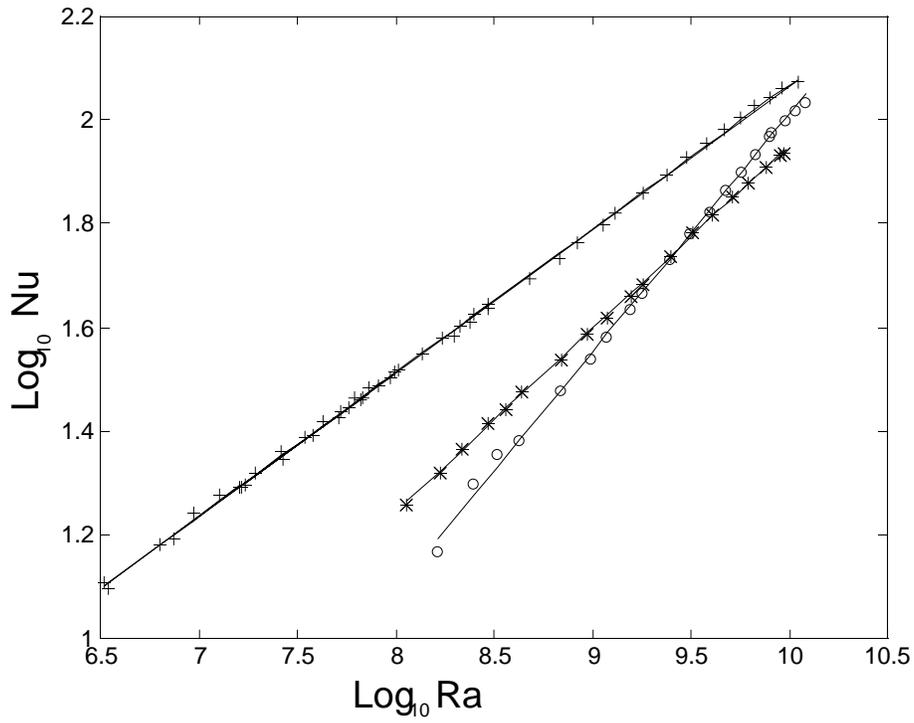}}
\caption{ Dependence of $Nu$  on $Ra$:
  with smooth bottom plate (+)  and with a  rough bottom plate in
 experiment III (*) with $\xi=2$    and in  experiment IV (o) with $\xi=1$. .
 In experiments III and IV $\lambda$ was always smaller than
$h_N$.
}
\end{figure}

\begin{figure}
\centerline{\epsfysize=0.7\linewidth \epsffile{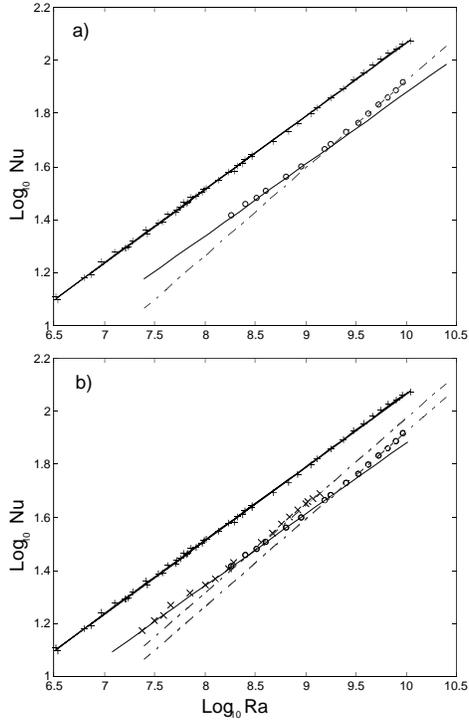}}
\caption{
  Dependence of $Nu$  on $Ra$:
  with smooth bottom plate (+) and  with a  rough bottom plate
in experiment I(o) and  in experiment II (x).
The roughness exponent was $\xi=2$.
a) The boundary layer thickness is  smaller than $h_N$ at $Ra> 10^9$
There is a clear change of slope when $\lambda < h_N$.
b) The results of experiment I are compared with those of experiment II.
For experiment II, $\lambda < h_N$ for $Ra > 3 \ 10^8$ which corresponds to
the transition point .
}
\end{figure}

\begin{figure}
\begin{center}
\centerline{\epsfysize=0.7\linewidth \epsffile{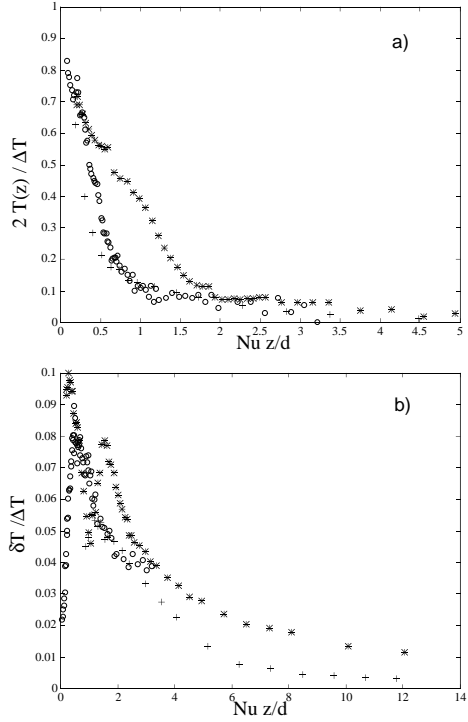}}
\caption{ Vertical  profiles of the mean temperature (a) and of
the rms of the temperature fluctuations (b). The symbol (+)
corresponds to measurements done with a smooth plate whereas (o)
and (*) correspond to the rough case with $\lambda > h_N$ (o)
($Ra=3 \ 10^8$) and   $\lambda  < h_N$ (*) ($Ra=10^{10}$). }
\end{center}
\end{figure}

\begin{figure}
\centerline{\epsfysize=0.7\linewidth \epsffile{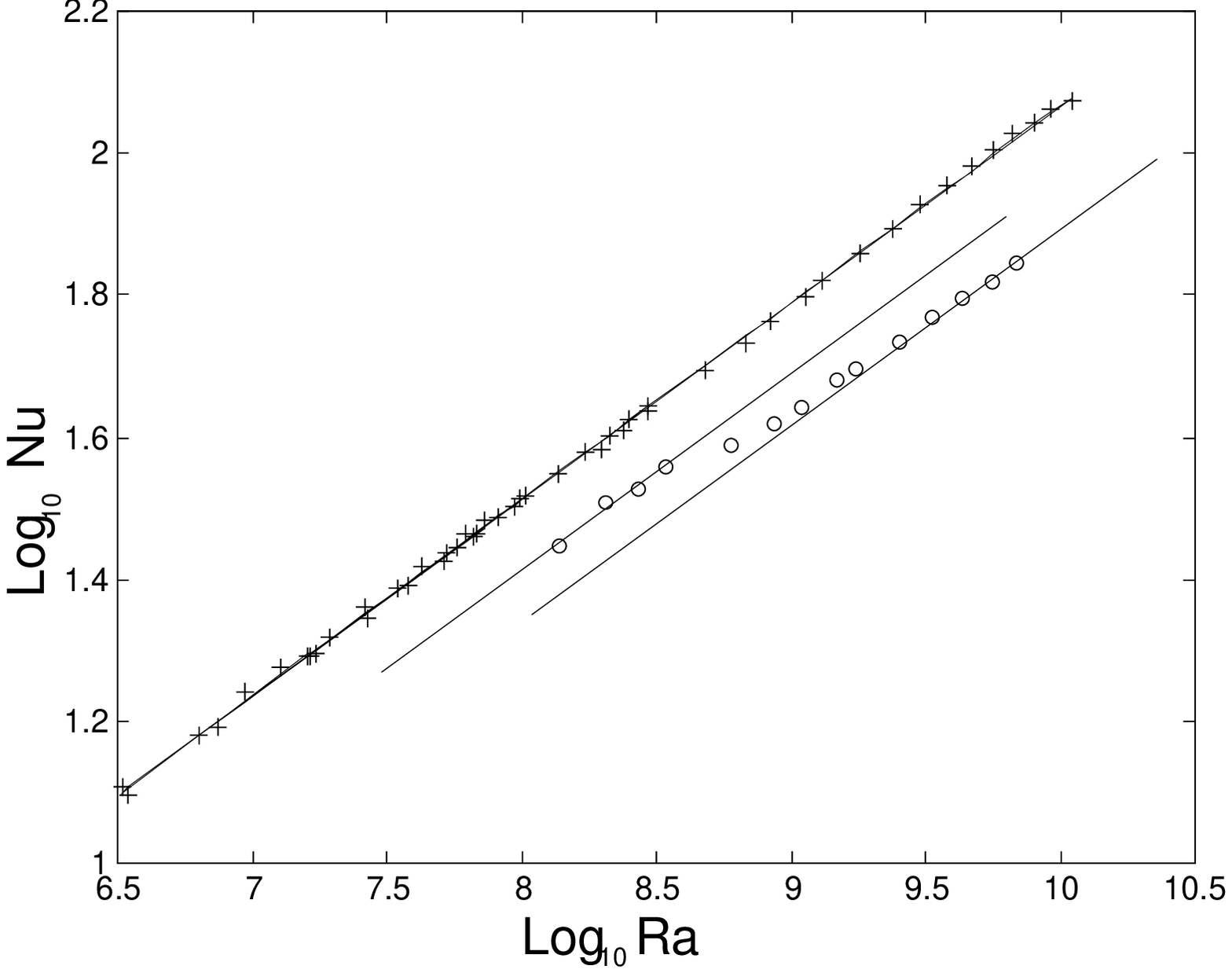}}
\caption{ Dependence of $Nu$  on $Ra$:
 with smooth bottom plate (+)  and  with a  rough bottom plate
 in experiment V  (o).
The roughness has just one size $h_N=h_1=2mm$ in this case.
 }
\end{figure}


\begin{thebibliography}{99}

\bibitem{SigII} E. Siggia, Annu. Rev. Fluid Mech. (1993).


\bibitem{Cast} B. Castaing, G. Gunaratne, F. Heslot, L. Kadanoff, A.
Libchaber,
 S. Thomae, X. Wu, S. Zalesky., G. Zanetti, J. Fluid Mech. 204, 1 (1989).

\bibitem{Chilla} F. Chill\'a, S. Ciliberto, C. Innocenti, E. Pampaloni,
Nuovo
 Cimento D 15, 1229 (1993);
 F. Chill\'a, S. Ciliberto, C. Innocenti, Europhys.
  Lett. 22, 681 (1993).

\bibitem{Chav} X. Chavanne, F. Chill\'a, B. Castaing, B. Hebral,
B. Chabaud, J. Chaussy, Phys. Rev. Lett. 79, 3648 (1997).


\bibitem{Siggia}  E. Siggia, B. Shraiman, Phys. Rev. A, 42, 3651 (1990).

\bibitem{Lohse}  S Grossmann, D. Lohse, preprint

\bibitem{Toschi} R.Benzi, F. Toschi, L. Tripiccione, chao-dyn/9808020



\bibitem{HWK} L.N.Howart, R. Krishnamurty, J. Fluid Mech., 170, 385
(1986);
 R. Krishnamurty, L.N.Howart, Proc. Natl. Acad. Sci. USA 78, 1981 (1981).

\bibitem{Sano} M. Sano, X. Z. Wu, A. Libchaber, Phys. Rev. A, 40, 6421
(1989).

\bibitem{Xia} K.Q.Xia, S.L. Lui, Phys. Rev. Lett., 79, 5006  (1997).

\bibitem{Gollub} T.H. Solomon, J. P. Gollub, Phys. Rev. A, 43, 6683 (1991).

\bibitem{Cili}  S. Ciliberto, S. Cioni, C. Laroche, Phys. Rev. E 54, R5901
(1996).


 \bibitem{Benzi} F. Toschi, R. Benzi, to be published.

\bibitem{Shen} X. Shen, P. Tong, K.Q.Xia, Phys. Rev. Lett. 76, 908 (1996).

\bibitem{Du} Y. B. Du, P. Tong, Phys. Rev. Lett. 81, 987 (1998).


\bibitem{Schl} H. Schlichting, {\it Boundary Layer Theory},
(McGraw-Hill, New York 1968).

\bibitem{Morn} N. Mordant, J.F.Pinton, F. Chill\'a, Journal de Physique 2,
VII, 1729 (1997), and reference therein.




 \bibitem{Vill} E. Villermaux, Phys. Rev. Lett. 81, 4859 (1998).





\end{thebibliography}
 \end{document}